# Electric power generation of a liquid self-assembled drop on a semiconductor surface

Przemysław Korasiak[1], Mateusz Pławecki[2], Edward Rówiński[2]

[1]*Institute of Automation, Opole University of Technology, Prószkowska Street 76, 45-758 Opole, Poland*
[2]*Institute of Materials Science, University of Silesia, 75 Pułku Piechoty Street 1A, 41-500 Chorzów, Poland.*

**Abstract.** The technological innovation of the direct conversion of solar energy to electric plays an important role in electric power generation. Earlier discussions of band bending in a semiconductor contacting a metal and liquid electrolyte solutions containing redox couples with different electrochemical potentials should not overshadow the fact that under absorption of photons takes place in a solar cell, which can generate free charge for an electrical circuit. Here we propose new band bending of ZnO and $Cu_2O$ semiconductors induced by a liquid self-assembled microdrop of a physiological salt solution. The drop/semiconductor interface under incident light irradiation of 0.1 $Wcm^{-2}$ exhibits a photovoltaic effect with efficiency of 19–24 % and enhanced electric power generation with electron or hole transport. We provide experimental and theoretical evidence that the architecture of the liquid drop/semiconductor interface will enable it to act as an electric power generator system in the near future.

**Introduction**

Partial wetting of a liquid on a smooth substrate is a general problem in science and engineering.[1–10] These problems derive from Young's equation.[2,7,10] The line energy associated with the triple-phase contact line is a function of local surface defects (chemical and topographical).[9,10] Contact angles can be divided into static and dynamic angles.[10] A static contact angle is measured when a drop is positioned on a solid surface and the three-phase boundary does not move. When the three-phase boundary moves, dynamic contact angles can be measured; these are referred to as advancing and receding angles. From this concept, we can back-calculate the Young equilibrium contact angle as a function of the maximal advancing and minimal receding contact angles. The liquid/solid interfaces are much harder to probe than the corresponding solid/vacuum interfaces; essentially, all experiments making use of electron beams are rendered unsuitable when a fluid is present.[10] Recently, electric voltage has been generated by spin-current generation from a fluid motion, i.e. spin hydrodynamic generation.[11-31-34] Controlling the motion of liquid drops on solid surfaces is crucial in many natural phenomena.[12–14] These phenomena makes mechanical spin-current and electric generators.[14]

However, Fermi level pinning refers to a situation involving band bending in a semiconductor contacting a liquid electrolyte solution containing redox couples with different electrochemical potentials[15]. Illumination of a liquid/semiconductor interface can result in an photovoltage output independent of the solution's potential.[15] This means that the liquid/semiconductor interface plays an important role in the quest for clean renewable energy and that it is possible to manipulate single molecules, atoms, and even electrons.[15–19] To the best of our knowledge, there have not yet been any reports of the photovoltaic effect in a liquid microdrop/semiconductor interface without the use of a metal electrode in a microdrop containing physiological salt solution.

Here we demonstrate that a liquid self-assembled 5-μl drop of physiological salt solution on a semiconductor (zinc oxide, ZnO; cuprous (I) oxide, $Cu_2O$) surface under incident light irradiation exhibits a photovoltaic effect and enhanced electric power generation. A new measurement method permits the elimination of the metallic electrode in the liquid microdrop; the resistor is used as a load to measure time evolution of both the photocurrent and photovoltage. It is important to note that the resistor's calibration value is specific for the output electric power. Our main observations show that the interface between the microdrop of physiological salt solution and the semiconductor surface is possible as long as (1) the microdrop is in contact with the semiconductor surface; (2) the time-dependent contact line of photoelectric quantities is pinned to its sample reference; (3) the water evaporates.

**Results**

**Photovoltaic performances.** To obtain better insight into the mechanism of the photovoltaic effect in a liquid microdrop/semiconductor interface, we have performed detailed electric investigations on interfaces of a 5-µl drop of physiological salt solution with ZnO and with $Cu_2O$ with fixed irradiations of 250 W/m$^2$ and 500 W/m$^2$ and fixed temperatures of 50°C and 70°C, respectively. Figures 1 and 2 show several interesting things.

To illustrate, the band bending (barrier height) of the semiconductor could be applied to the electric measurement method without using a metal electrode in the physiological salt solution through irradiation of the interface, using the resistor as a load to measure the time evolution of both the photocurrent and photovoltage. Figure 1a shows a schematic diagram of the experiment. There are two fundamental top contacts. The first is between the metallic electrode and the semiconductor surface; the second is between the liquid microdrop and the same semiconductor surface as in the first (top) contact. The semiconductors are ZnO and $Cu_2O$ layers. The substrates are a silicon p-n junction and $Cu_2O$/Cu(100) and $Cu_2O$/polycrystalline copper interfaces. These substrates are designated as the reference sample. The top electrodes are made of polycrystalline pure silver, whereas the bottom electrodes are made of pure silver, polycrystalline copper (poly-Cu), and crystalline copper (Cu(100)).

For these reasons, both electrodes and reference samples build the layered sandwich-type architecture: ZnO/Ag/textured layer of n-type silicon/silicon p-n junction/crystal layer of Si(100)/Al/Ag, Ag/$Cu_2O$/poly-Cu, and Ag/$Cu_2O$/Cu(100) solar cells. The microdrop contact can be made by simply avoiding contact between the top electrode and the semiconductor. However, the microdrop in such top-contact semiconductor surfaces acts as an additional interface. In this paper, a new, additional interface was developed and used in PV devices.

Physiological salt in solution displays a broadened peak due to the absorption of higher photon energy in the range 2.7–3.3 eV. This absorption peak plays a dominant role in the photovoltaic effect.[20–22] Figures. 1c–e and 2 illustrate the time evolution of electric power over load, photocurrent, and photovoltage for the drop/semiconductor interfaces. From all shapes of the curves, we see that, immediately following the partial wetting of drops on the semiconductor surface, they change their shapes very quickly. However, the power is also very sensitive to changes in the polycrystalline and crystalline coppers (Fig. 2). It should be noted that the quick jumps are due to the instability of band bending in the semiconductor as suggested by the photovoltage plots (Figs. 1e, 2c). Next, the band bending at equilibrium is pinned to around 0.02 eV for the n-type of ZnO semiconductor or 0.00005 eV for the p-type of $Cu_2O$ semiconductor. At the end of the evaporation process, the exponential behaviour of power relaxes back to equilibrium (Figs. 1e, 2c). This moment is accompanied by the formation of salt crystals on various semiconductor surfaces.[23-25]

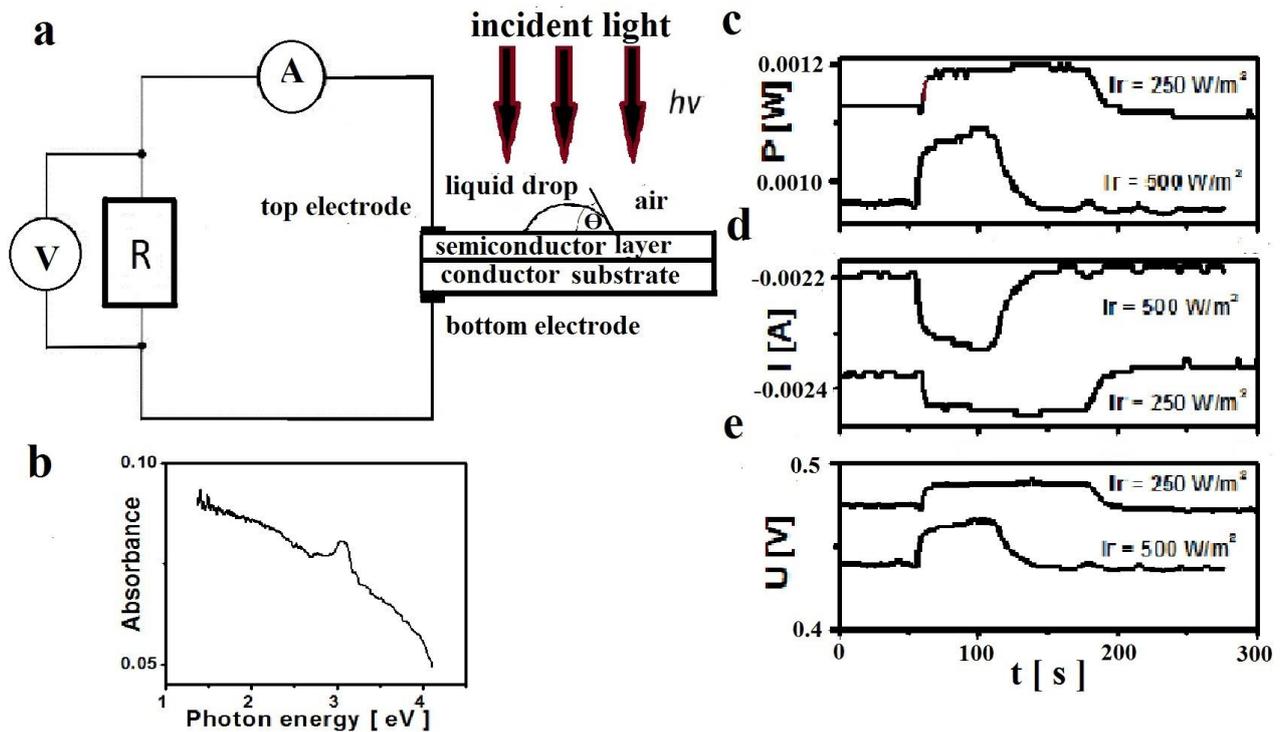

**Figure 1.** Schematic diagram of the experiment, absorption spectrum, and time evolution of the electrical quantities of ZnO layer in contact with the 5-µl physiological salt drop during fixed irradiations (fixed temperatures) Ir=250 W/m$^2$ (50°C – top curve) and Ir=500 W/m$^2$ (70°C – bottom curve), respectively; **a)** the experimental setup. A and V denote ampere meter and voltage meter, respectively. R is the external resistor; the source of photons is denoted as hν; a simple consists of the liquid drop, the semiconductor layer, the conductor substrate, and top and bottom electrodes; **b)** the absorption spectrum of the physiological salt solution in a photon energy range of 1.38–4.1 eV; **c)** time-dependent power of the physiological salt drop/ZnO layer interface; **d)** time-dependent photocurrent of the liquid drop/ZnO layer interface; **e)** time-dependent photovoltage of the liquid drop/ZnO layer interface.

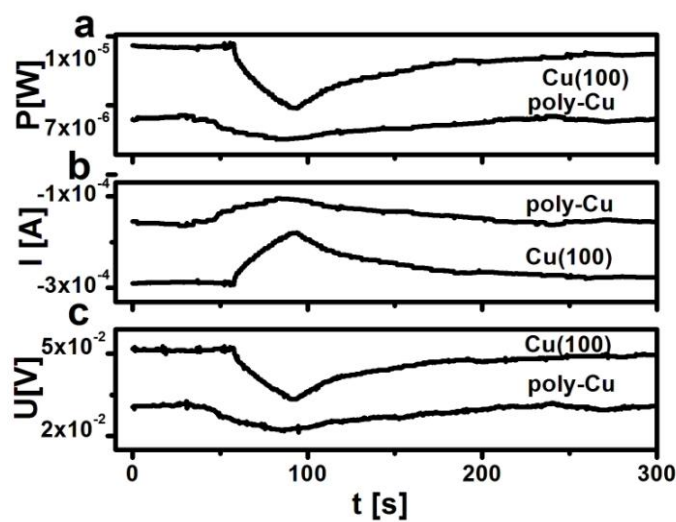

**Figure 2.** Time evolution of the electrical quantities of polycrystalline Cu$_2$O layers during isothermal photovoltaics (temperature 50°C and irradiation 250 W/m$^2$) in contact with a 5-µl-volume liquid drop of physiological salt solution; **a)** time-dependent power of the physiological salt drop/Cu$_2$O layer interface; **b)** time-dependent photocurrent of the liquid drop/Cu$_2$O layer interface; **c)** time-dependent photovoltage of the liquid drop/Cu$_2$O layer interface.

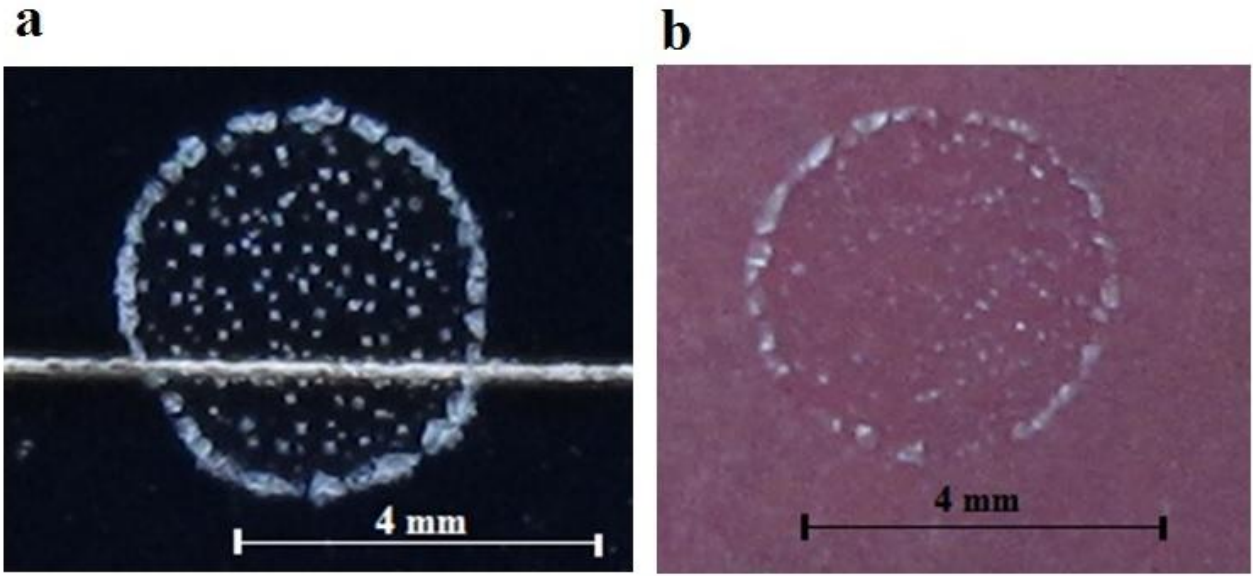

**Figure 3. Photograph of an evaporating 5-ul drop of physiological salt solution on ZnO and Cu$_2$O layers (temperature 50°C and irradiation 250 W/m$^2$; a)** ring effect of crystallising solution droplet evaporation on the ZnO layer; **b)** ring effect of crystallising solution droplet evaporation on the Cu$_2$O layer.

The crystal deposits left after complete drying for physiological salts show that the final crystallisation patterns are very similar (Fig. 3). Systematically, NaCl crystallisation leads to a ring-like deposit with a large number of tiny crystals. For this reason, the salts crystallise and grow during the drying process, which is independent of the substrate.

In the nonequilibrium case, the 500 W/m irradiation of the drop takes place rapidly; its temperature is close to the boiling temperature of water. The evaporation process is more intensive. Taken together, the results suggest that the photovoltaic effect of the interface can result in an output of electrical power which is dependent on the type of semiconductor.

**Description and application of a technique to measure the initial contact angle of partial wetting of a physiological salt solution on ZnO and Cu$_2$O surfaces.** The volume of liquid drop at a given moment, $V(t)$, is:[26]

$$V(t) \cong \frac{\pi \cdot R^3 \theta_c}{4}(1 - \frac{t}{t_f}) \qquad (1)$$

where $R$ – the radius of drop, $t$ –time, $\theta_c$ –initial contact angle, $t_f$ – effective total drying time. Equation (1) predicts that the initial contact angle is a quantitative measure of the partial wetting of a semiconductor (solid) by a liquid microdrop. In the case of $t_f \gg t$, the average initial contact angle for the 32 samples was found to be 54°. In this context, the ZnO and Cu$_2$O semiconductors are considered hydrophilic surfaces.

**Simulation studies of current-voltage characteristics.** In the experimental results, there is clearly a depletion layer adjacent to the semiconductor-drop interface, where the redox energy is close neither to the bottom of the conduction band nor to the top of the valence band. The barrier height between the semiconductor materials and the drop may be formed by a van der Waals force; the semiconductor bands start out bent due to surface states.

Briefly, for any individual case, the barrier uses a proportional model to establish the relationship between the potential redox of the physiological salt solution and the electronic structure of the semiconductor.[4,19,27–30]

We consider two different barrier heights: 1) $E_n^{SBH}$ is the barrier height against electron flow between the drop and the n-type semiconductor; 2) $E_p^{SBH}$ is the barrier height against hole flow between the drop and the p-type semiconductor. The redox potential of physiological salt is $-0.8$ V with the value of the Fermi level, $E_{F,redox}$, equal to $-5.3$ eV. $E_{F,redox}$ is situated between the top energy of the valence band ($E_{VB}$) and the bottom energy of the conductor band ($E_{CB}$). Researchers often use the Shockley equation of the current-voltage characteristics of a solar cell.[19,27,28,31-34] Here, we propose a modified Shockley equation; the total current, $I$, is then:

$$I \cong S \cdot A^* \cdot T^\alpha \cdot \exp[\frac{qV}{n^{eff}} - E_i^{SBH})/(k_B T)] + \frac{V - I \cdot R_s}{R_{sh}} - I_{sc} \qquad \text{for} \quad qV \gg k_B T \qquad (2)$$

$$I_{sc} = S \cdot Q \cdot (1-r) \cdot [1 - Exp(-\alpha \cdot d)] \cdot q \cdot n_{ph}(E_g) \qquad (3)$$

$$E_n^{SBH} \cong \zeta \cdot \left(\frac{E_{CB} + E_{F,redox}}{2}\right) - \chi_s \qquad (4)$$

$$E_p^{SBH} \cong E_s - \zeta \cdot \left(\frac{E_{VB} + E_{F,redox}}{2}\right) \qquad (5)$$

where $I$ – the current, $V$ – the voltage, $S$ – the contact area between the liquid microdrop and semiconductor, $A^*$ – Richardson's constant, $T$ – the absolute temperature, $\alpha$ – the constant, $q$ – the elementary electric charge, $k_B$ – the Boltzmann constant, $n^{eff}$ – the effective ideality factor, $E_i^{SBH}$ – the Schottky barrier height, $i = n, \ p$, $n$ – the n-type semiconductor, $p$ – the p-type semiconductor, $R_{sh}$ – the shunt resistance, $R_s$ – the series resistance, $I_{sc}$ – the short-circuit current, $Q$ – collection efficiency, defined as the ratio of the carriers passing through the circuit to those generated in the layer, $r$ – the reflection coefficient, $\alpha$ – the absorption constant, $d$ – the thickness of the absorbing semiconductor layer, $E_g$ – the band-gap energy of the absorbing semiconductor, $n_{ph}(E_g)$ – the number of photons per second per unit area of a single junction whose energy is great enough to generate hole-electron pairs in the semiconductor layer, $\zeta$ the dimensionless parameter, $\chi_s$ – the electron affinity of the semiconductor, $E_s$ – the work function of the semiconductor.

The theory describes current-voltage characteristics of a liquid drop/semiconductor interface of a solar cell. Figure 4 presents theoretical current-voltage characteristics for a solar cell and its electric power. Each square centimetre of area can produce continuous power of ~0.019 and ~0.024 W for drop/ZnO and drop/Cu$_2$O solar cells, respectively. In our approach, the power conversion efficiency ($\eta$) of a liquid microdrop/semiconductor solar cell is defined by the well-known equation[31,33]:

$$\eta = \frac{P_{max}}{S \cdot I_r} \cdot 100\% \qquad (6)$$

where $P_{max}$ – the maximum power, $I_r$ – the irradiation (expressed in W/m$^2$). Theoretical values of power conversion efficiencies are ~19 % and ~24 % for the drop/ZnO and the drop/Cu$_2$O solar cells, respectively. The high value of efficiency suggests that the photovoltaic effect of the interface produces electric power generation which is dependent on the type of semiconductor, band-gap energy, and Fermi level of the redox system.

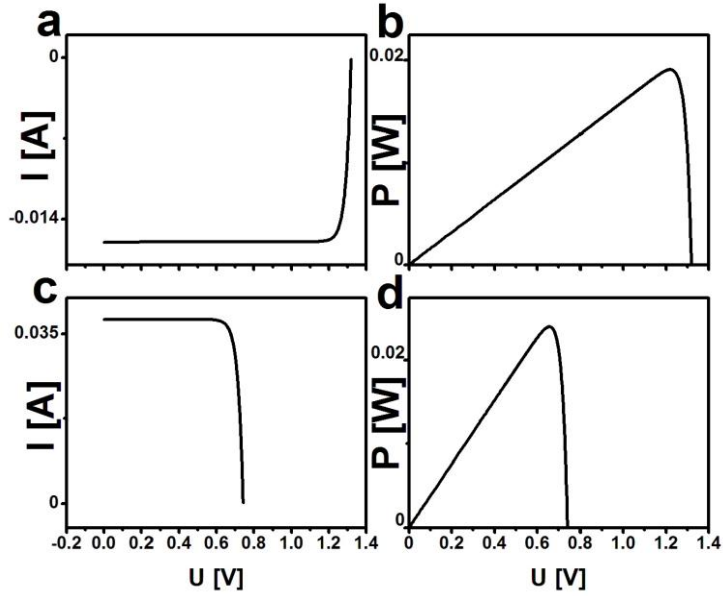

**Figure 4. Theoretical current-voltage characteristics and power-voltage curves of liquid drop of physiological salt solution/semiconductor solar cells at room temperature and irradiation 0.100 W/cm² obtained by equations (2)–(5); a)** theoretical current-voltage characteristics of the liquid drop/ZnO semiconductor solar cell. The parameters are $S = 1\,\text{cm}^2$, $A^* = 120\,\text{AK}^{-2}\text{cm}^{-2}$, $T = 291\,\text{K}$, $n^{\text{eff}} = 1$, $e = 1.6 \times 10^{-19}\,\text{C}$, $E_n^{SBH} = 1.7\,\text{eV}$, $k_B T = 0.026$ eV, $R_{sh} = 40\,\text{k}\Omega$, $R_s = 50\,\Omega$, $Q = 1$, $(1 - \exp(-\alpha \cdot d)) = 0.63$, $n = 2.67$, $r = 0.21$, $E_g = 3.4\,\text{eV}$, and $n_{ph}(E_g) = 2 \times 10^{17}\,\text{s}^{-1}\text{cm}^{-2}$; **b)** power-voltage curve for a liquid drop/ZnO semiconductor solar cell; **c)** theoretical current-voltage characteristics of a liquid drop/Cu₂O semiconductor solar cell. The parameters are $S = 1\,\text{cm}^2$, $A^* = 120\,\text{AK}^{-2}\text{cm}^{-2}$, $T = 291\,\text{K}$, $n^{\text{eff}} = 1$, $e = 1.6 \times 10^{-19}\,\text{C}$, $E_n^{SBH} = 1.1\,\text{eV}$, $k_B T = 0.026$ eV, $R_{sh} = 40\,\text{k}\Omega$, $R_s = 50\,\Omega$, $Q = 1$, $(1 - \exp(-\alpha \cdot d)) = 0.63$, $n = 2.97$, $r = 0.25$, $E_g = 2.2\,\text{eV}$, and $n_{ph}(E_g) = 5 \times 10^{17}\,\text{s}^{-1}\text{cm}^{-2}$; **d)** power-voltage curve for a liquid drop/Cu₂O semiconductor solar cell.

**Discussion**

We have presented the first measurements that follow the time evolution of photocurrent and photovoltage in a liquid self-assembled drop on a semiconductor surface at both a fixed level of irradiation and a fixed temperature. Our study reveals the electric power generation of the liquid drop/semiconductor interface with electron or hole transport. Our electrical measurement method enables the time evaluation of electrical quantities in the drop-semiconductor contact line, a significant barrier that heretofore has limited by the necessity of using a metal electrode in liquid. Power is also very sensitive to changes in the substrate of polycrystalline and crystalline coppers. For example, we provide experimental evidence that the architecture of a 5-μl drop of physiological salt solution/ZnO interface produces electric energy of ~0.012 J. The electric energy depends linearly on the volume of the drop, given by E = 0.0024*10⁶ * V_V, where E – the electric energy (expressed in J), V_V – the volume of the liquid drop (in l). The photovoltaic effect vanishes following the evaporation of water.

Our simulation studies, based on equations (2)–(5), can be directly generalised to describe the photovoltaic effects of interfaces; such systems enable even greater values of efficiency (~19–24 %). The results predict the band bending of a semiconductor induced by a liquid self-assembled drop. We discovered that the liquid microdrop/semiconductor interface can transform the conversion of solar to electric energy at an extremely low cost. Scientific research has confirmed the diverse photovoltaic effects of the liquid drop/semiconductor interface and established its ability to act as an electric power generator system in the near future.

## Methods

**Materials and device preparation.** The $Cu_2O$ layer was deposited on crystalline and polycrystalline copper. The copper crystals were grown from 99.999 % Cu using the Bridgman technique, homogenised under a vacuum of $10^{-6}$ Torr at 1273 K for 72 h and cooled to room temperature at a rate of 6 K/h. All crystals were of the dimensions $1 \times 1 \times 0.5$ cm$^3$ with faces parallel to (100) (. Polycrystalline copper films were 0.5 mm thick and $2 \times 2$ cm$^2$ in area. The surface of the copper substrates was treated with chemical etching for about 15 s in an $HNO_3$ solution, followed by a second etching in HCl (50 %) for 15 s. The surface was then dipped in de-ionised water and dried in an Ar flux.

A very simple apparatus was used for the electrodeposition of $Cu_2O$ layers on copper surfaces, consisting of a thermostat, a glass holding the solution, two electrodes (a cathode and anode), and a standard electrical circuit device for electrolysis. The $Cu_2O$ was prepared similarly to the previously published method[5]; in brief, copper (II) sulphate ($CuSO_4$, 0.4 mol L$^{-1}$, Wako 97.5 %) and L-lactic acid (3 mol L$^{-1}$, Wako) were dissolved in distilled water. The pH of the electrolyte was adjusted to 12.5 by adding NaOH. The electrolyte temperature was kept at 60°C during electrodeposition. The current density of electrodeposition was carried out at 1.5 mA/cm$^2$ for 20 min. Faraday's law was applied to, the estimated thickness of the $Cu_2O$ layer. The $Cu_2O$ layer was about 1 μm with a roughness of 92.5 nm. The crystallite sizes of Cu and $Cu_2O$ were determined at 49.2(3) nm and 69.8(6) nm, respectively. A metal electrode of Ag was painted on top of the $Cu_2O$ layer using silver paste (TAAB, S270).

Monocrystalline solar cells were produced from silicon wafer substrates cut from column ingots grown using the Czochralski process. Crystalline silicon solar cells were characterised by highly phosphorous-doped $n_+$ (electron-producing) regions on the front surface of boron-doped p-type (electron-accepting) substrates to form p-n junctions. Back-surface $p_+$ field regions were formed on the back surface of the silicon substrate to suppress recombination of minority carriers (photogenerated electrons). The top electrode consisted of silver gridlines connected by a bus bar to form a double comb-shaped structure. The bottom electrode was formed by two series of silver stripes and polycrystalline aluminium. The front surface of the silicon solar cell was covered with micrometre-sized pyramid structures (textured surface). An anti-reflection coating of zinc oxide (ZnO, thickness ~10 nm) was overlaid on the textured silicon surface. The surface roughness was found to be ~0.2 nm. The average thickness of silicon cells was 250(9) μm, with an active area of $101 \times 11$ mm$^2$ and efficiency of ~15 %.

Liquid self-assembled drops of physiological salt solution were placed on the zinc oxide (ZnO) and cuprous (I) oxide ($Cu_2O$) semiconductor surfaces. The drops were placed on the *n*- and *p*-type semiconductor surfaces using a volume-adjustable micropipette. The volume of each drop was 5 μl at an impact velocity of 1 mm s$^{-1}$. Photographs of the evaporating 5-ul drop of physiological salt solution on ZnO and $Cu_2O$ layers obtained by Canon G11 10 MP Digital Camera. We studied 32 samples.

**Material characterisation.** The structural studies of polycrystalline Cu and $Cu_2O$ materials were carried out by means of XRD and GIXD on an Empyrean PANalytical powder diffractometer using CuK$\alpha_1$ radiation. Grazing incidence geometry with an incident angle of 1° was applied. The backscatter Laue diffraction pattern of Cu(100) and Si(100) crystals was produced on an XRT (X-ray topography) 100 CCM diffractometer manufactured by EFG GmbH. The AFM image and micrograph were determined by a Hysitron TI 950 TriboIndenter equipped with a Q-Scope 250 and JEOL-JSM-6480 scanning electron microscope, respectively. Characteristics of materials are shown in Supplementary Figures S1-S3.

Measurements of emission and excitation spectra were recorded using a FluoroMax-4 spectrofluorometer (Horiba, USA) equipped with automated polarisers. The light source was a 150-W ozone-free xenon arc lamp blazed at 330 nm (excitation) and 500 nm (emission). The characterization

Both current-time and voltage-time curves were carried out using a 2400 Series SourceMeter (Keithley Instruments) using a 200-ohm load resistor. The source was a 75-W wolfram lamp equipped with a sunlight filter to match the emission spectrum of the lamp to 250 and 500 W/m$^2$.

# Supplementary Figures

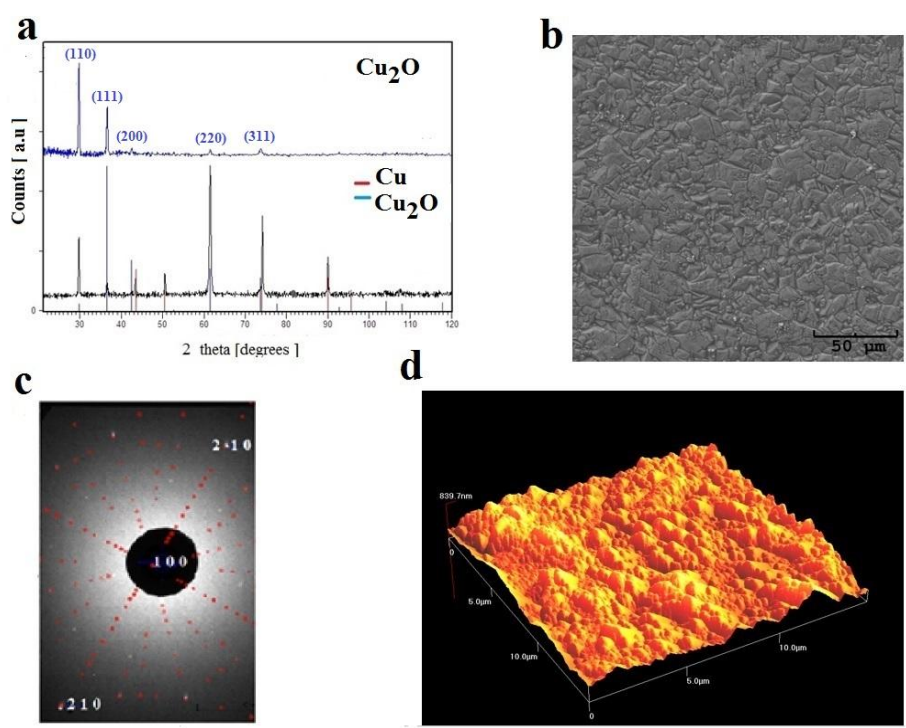

**Supplementary Figure S1. Characteristics of a $Cu_2O$ layer deposited on the polycrystalline and crystalline copper substrates;** a) GIXD at $1^o$ incidence angle of the $Cu_2O$ layer and XRD patterns of $Cu_2O$ layer and polycrystalline copper substrate; b) micrograph of $Cu_2O$ layer; c) backscatter Laue diffraction pattern from a Cu(100) crystal; d) AFM image of a 4-nm-thick layer of $Cu_2O$ deposited on Cu(100) substrate.

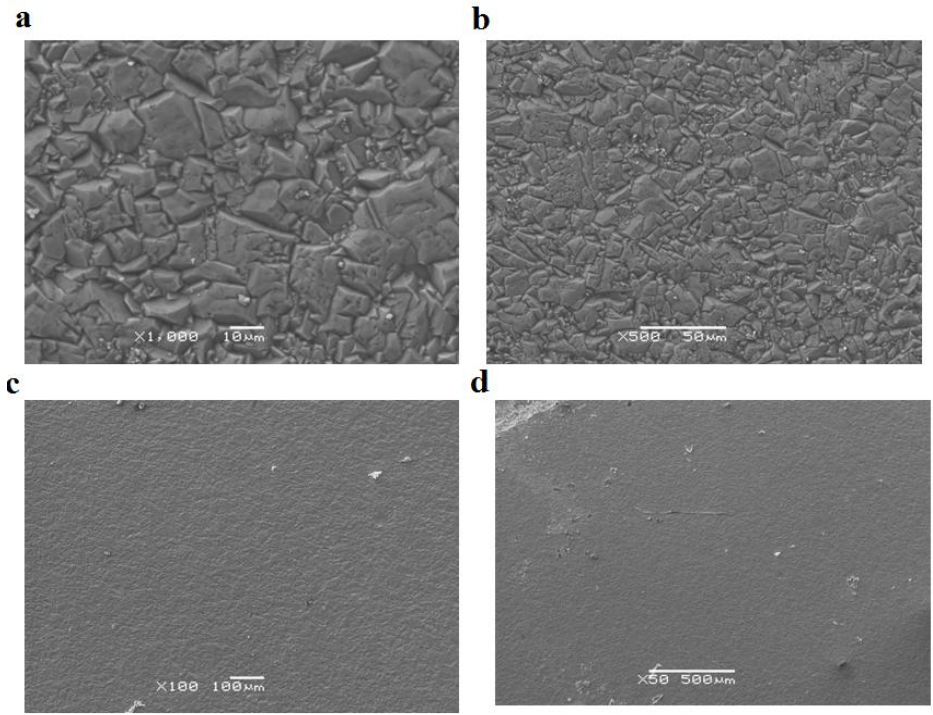

**Supplementary Figure S2.** Micrographs of a $Cu_2O$ layer obtained from a scanning electron microscope at magnifications : a) x1000; b) x500; c) x100; d) x50.

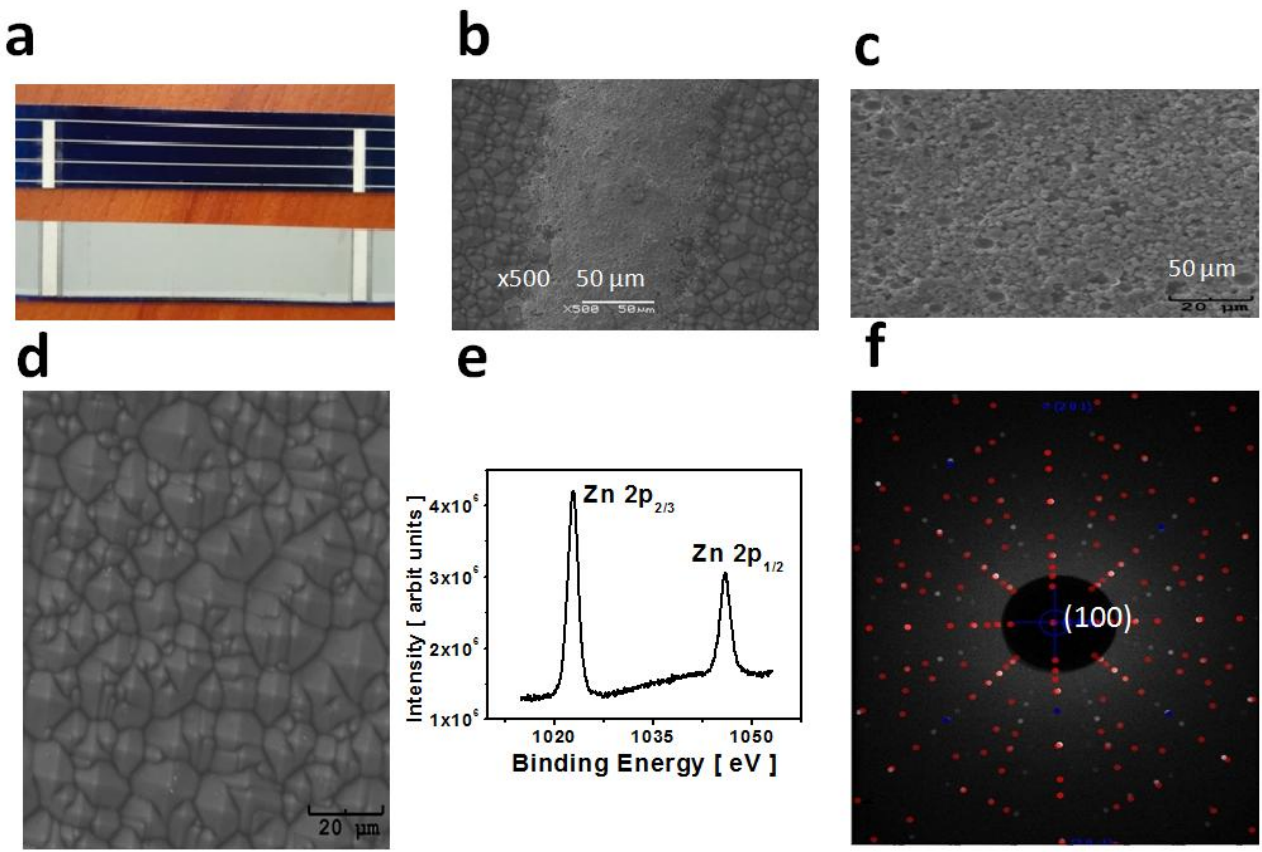

**Supplementary Figure S3. Characteristics of silicon solar cell;** a) photographs of top and bottom solar cell; b) micrograph of silver electrode; c) micrograph of aluminium bottom electrode; d) micrograph of type-*n* semiconductor of textured silicon layer covered by ZnO antireflection thin layer; e) X-ray photoelectron core lines of Zn2p 3/ 2 and Zn2p1/ 2 of ZnO layer; f) backscatter Laue diffraction pattern from a Si(100) crystal.